\documentclass[final, nomarks]{dmtcs-episciences}

\usepackage{etex}
\usepackage{amsmath}
\usepackage{amssymb,amsthm,url}
\usepackage{stmaryrd}
\usepackage{doi}

\newtheorem{open}{Open problem}
\newtheorem{definition}{Definition}

\usepackage[numbers]{natbib}

\title{Introduction to local certification}
\author{Laurent Feuilloley}

\affiliation{Université Lyon 1, LIRIS}%


\received{2020-04-14}

\accepted{2021-09-05}

\publicationdetails{23}{2021}{3}{9}{6280}

\begin{document}

\maketitle

\begin{abstract}

A distributed graph algorithm is an algorithm where every node determines its output by inspecting its local neighborhood. 
As distributed environments are subject to faults, an important issue is to be able to check that the output is correct, or in general that the network is in proper configuration with respect to some predicate. 
One would like this checking to be very local, to avoid using too much time. 
Unfortunately, most predicates cannot be checked this way, and that is where certification comes into play. 
Local certification (also known as proof-labeling schemes, locally checkable proofs, or distributed verification) consists in assigning labels to the nodes, that certify that the configuration is correct. 
In this paper, we present several different perspectives for studying local certification: as a part of self-stabilizing algorithms, as a labeling problem, or as a non-deterministic distributed decision.

This paper is an introduction to the domain of local certification, giving an overview of the history, the techniques and the current research directions.
\end{abstract}

\section{Introduction}

Let us consider an informal example that illustrates the concept of local certification. (Formal definitions will follow, and a more formal writing will be used in the rest of the paper.)
The scenario is the following.
The nodes of a connected graph have to decide collectively whether the graph they belong to is a path.
Every node wakes up, with no knowledge about the graph, inspects its local neighborhood, and then decides either to stay silent or to raise an alarm.
If the graph is a path, all nodes should stay silent, but if the graph is not a path, then at least one node should raise an alarm.
If a node stays silent, we will say that it accepts, and if it raises an alarm we will say that it rejects.

There are some cases where it is easy to detect that something is wrong: if a node sees that it has more than two neighbors, then it can safely raise an alarm.  
But what if the graph is a cycle?
Then every node sees that it has two neighbors, and as the graph is not a path, at least one node should reject.
If we assume that the nodes are all identical, then the only possibility is that all nodes reject.
Then consider a node in the middle of a long path.
It sees two neighbors, thus it has the same partial view of the network as the nodes in the cycle, and as a consequence, it rejects.
This is an incorrect behavior: a correct instance (a path) has a node raising an alarm.
Note that if instead of accepting paths and rejecting cycles, we aimed for the converse, that is accepting cycles and rejecting paths, the task would be much easier: we could simply make nodes of degree one reject.

\begin{figure}[!h]
\begin{center}
\begin{tabular}{ccc}
\hspace{-0.5cm}
\includegraphics[scale=0.78]{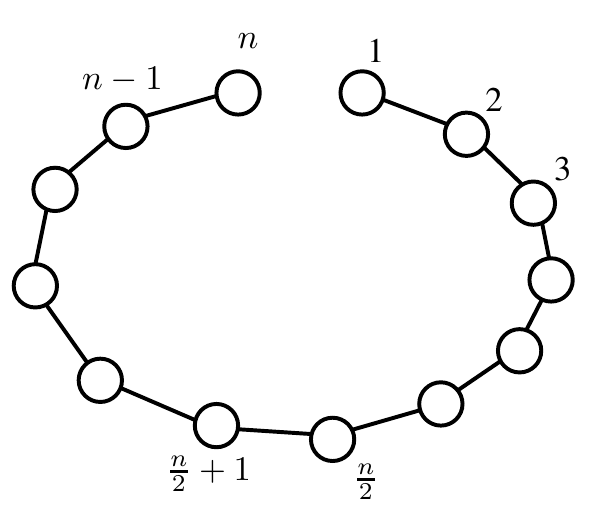}
&
\hspace{-0.5cm}
\includegraphics[scale=0.78]{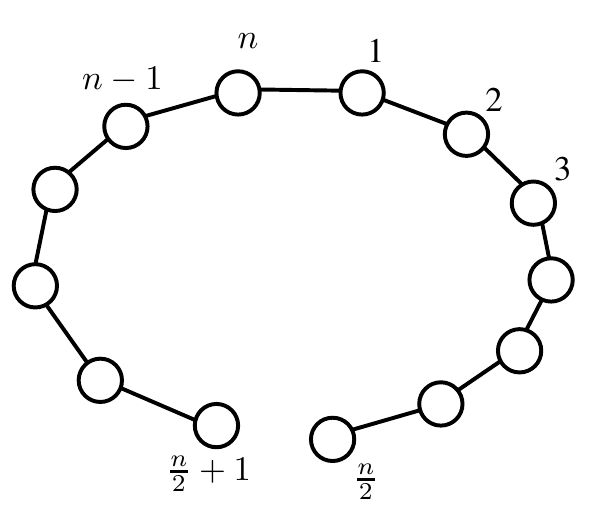}
&
\hspace{-0.5cm}
\includegraphics[scale=0.78]{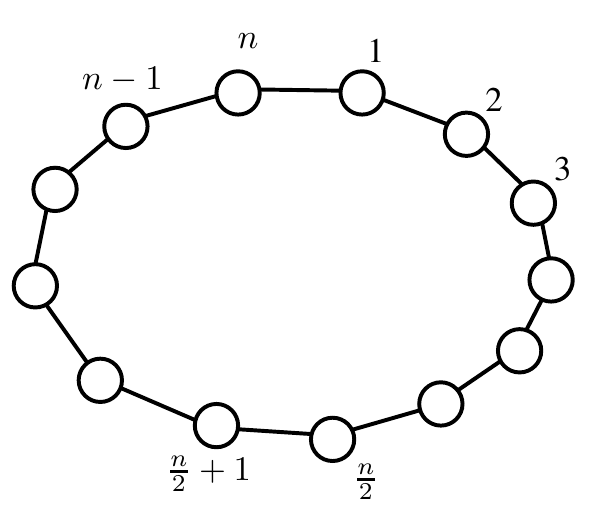}
\end{tabular}
\caption{\label{fig:limit-basic} This figure illustrates that, even with distinct identifiers, it is not possible to locally distinguish between paths and cycles.
We consider three graphs on $n$ nodes with unique identifiers between 1 and $n$. 
The two first graphs are paths, and the third is a cycle.
Consider the view of a node of the third graph, that is, the node itself and its two neighbors.
By construction, this view also appears in at least one of the two paths.
Now, because the third graph is a cycle, one node $v$ should reject.
There is a node with the same view in one of the paths, and this node should also reject, because it has the exact same view of the system. This is a contradiction, because in a path no node should reject.}
\end{center}
\end{figure}

\begin{figure}[!h]
\begin{tabular}{cccc}
\includegraphics[scale=0.65]{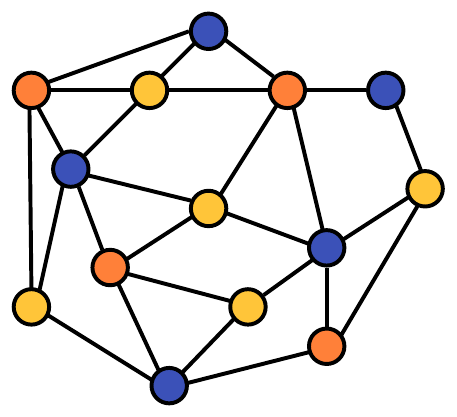}
&
\includegraphics[scale=0.65]{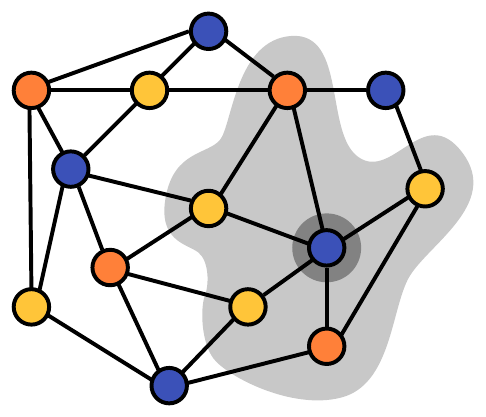}
&
\includegraphics[scale=0.65]{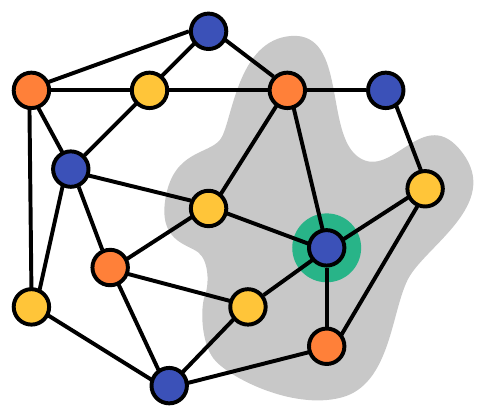}
&
\includegraphics[scale=0.65]{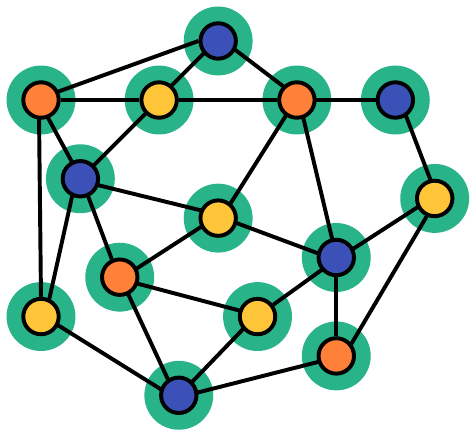}
\\
\includegraphics[scale=0.65]{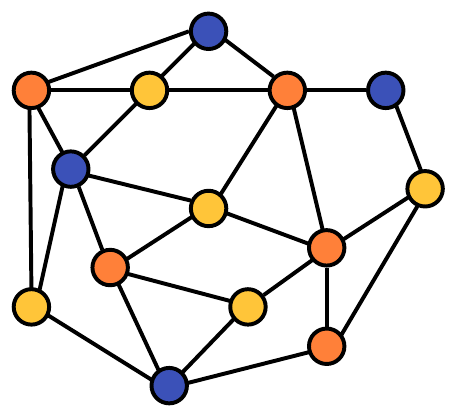}
&
\includegraphics[scale=0.65]{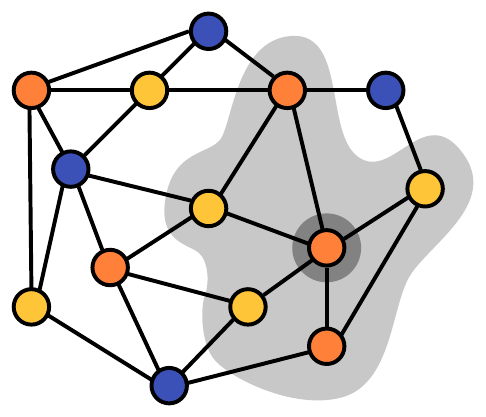}
&
\includegraphics[scale=0.65]{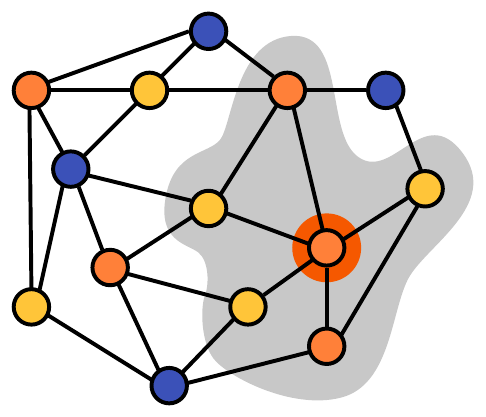}
&
\includegraphics[scale=0.65]{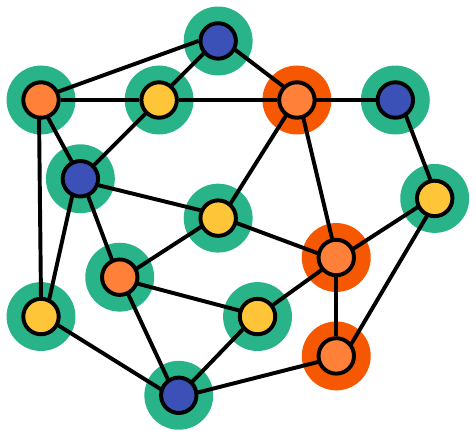}
\end{tabular}
\caption{\label{fig:example-coloring}
This figure illustrates that for some well-known property one can actually verify locally (without additional help).
The property is the following: every node of the graph is given a color as input, and no edge should have both of its endpoints of the same color.
The verification is very simple: every node checks that none of its neighbors has the same color as its own.
In the first row of pictures, the instance is correct, and every node accepts, but in the second row, several nodes notice that the coloring is not correct.
(In both rows, we highlight the behavior of one specific node, and then illustrate the collection of outputs.)
}
\end{figure}

We have just proved informally that, in our model, the nodes of a network cannot accept paths and reject cycles.
The assumption that the nodes are identical is actually not necessary for this impossibility result.
Indeed, if we assume that the nodes run the same algorithm but that they have distinct identifiers, which will be our standard model in this paper, the result still holds.
This is explained in Figure~\ref{fig:limit-basic}.
Some specific properties can actually be checked in the model above, for example the fact that a graph is properly 3-colored, as illustrated in Figure~\ref{fig:example-coloring}. 
But for most properties, this verification mechanism is too weak. The notion of certification originates from this limitation.
In order to verify a property, we could allow more resources, for example allowing the nodes to see further in the graph. 
But verification should not consume a lot of resources as it is not the main computation, but a secondary computation that is useful only in case of faults. Consequently, we want to keep the verification very local.
Another approach consists in giving some information to the nodes.
We will do the following: assign \emph{certificates} to the nodes that allow them to check the property.
For the sake of concreteness, let us go back to the path example.
How can we convince the nodes that they live in a path and not in a cycle?
One efficient way to do so is the following: chose one of the endpoints of the path, and give to every node its distance to this node~\cite{AfekKY90}. 
Now a node can see its distance and the ones of its neighbors. 
It can easily check that the distances are consistent: either (1) it has been assigned distance 0, and then it should have degree 1, or (2) it has a distance $d>0$, and it should check that one of its neighbors has distance $d-1$ and the other (if it exists) has distance $d+1$. 
The crucial point is that the nodes cannot be fooled. 
That is, if the graph is a cycle, and one assigns numbers to the nodes pretending that these are distances to one endpoint, at least one node will detect the inconsistency and reject (for example a node with the smallest distance). 
The labels assigned to the nodes are called the certificates, and the global mechanism is called local certification.
Now many questions can be asked. For example, the certification of paths we have just described uses labels on  $O(\log n)$ bits for graphs on $n$ nodes; is this optimal? 
For some other property, does there exist a similar mechanism? 
What if we allow the view of the nodes to be slightly larger? 
And also, this labeling given by an untrustworthy oracle is similar to NP in centralized computing, can we push the analogy further? This is what research in local certification is about, and that is the topic of this paper.
\paragraph*{Organization of the paper.}
The organization of the rest of the paper is the following. 
In Section~\ref{sec:history}, we give some historical perspective, describing where the notion of certification comes from and how it fits in the field of distributed graph algorithms. 
In Section~\ref{sec:definitions}, we introduce the definitions and vocabulary. 
In Section~\ref{sec:size}, we give an overview of what we know about the performance of local certification in terms of label size, both in terms of results and technique.
In Section~\ref{sec:complexity}, we take the more complexity-theoretic point of view of distributed decision.
Finally, in Section~\ref{sec:directions}, we describe the current research directions being explored, and provide a list of open problems. 
\section{Context and historical perspective}
\label{sec:history}
Local certification can be seen as part of two related lines of work in the theory of distributed computing. 
On the one hand, the study of locality, and on the other hand the study of self-stabilization.
Both of these areas are about construction tasks (\emph{e.g.} computing a proper coloring of the graph), and not about decision tasks like the ones we study in this paper. Nevertheless, decision tasks appear implicitly at theirs cores.
Let us give some key notions about these two fields before we move to certification itself.

\paragraph*{Study of locality and locally checkable labelings.}
The idea of locality in distributed computing consists in focusing on the time it takes for information to be disseminated in the network, putting aside (or at least on the background) other issues such as asynchrony, failures or bandwidth limitations. 
A classic model is the LOCAL model, where a node can see a neighborhood at distance $T$ in the network, and choose an output given this partial view. 
The main goal from the algorithm designer perspective is then to minimize $T$ for some given task. 
The study of locality in distributed computing dates back to the mid-80's (\emph{e.g.}~\cite{Luby86, AlonBI86, ColeV86, GoldbergP87}) and to 90s with the seminal papers \cite{Linial92} and~\cite{NaorS95}, and was developed into a book in 2000~\cite{Peleg00}.
In \cite{NaorS95}, motivated by~\cite{AfekKY90}, the authors define a class of (construction) problems called \emph{locally checkable labeling}, or LCL for short. 
For these problems, the nodes may have inputs, and they have to compute outputs. 
A problem is in LCL if there exists an algorithm that, given a view at some constant distance, with the inputs and outputs, can decide whether the solution computed is correct. 
This is exactly the kind of problem for which no certification is needed, and coloring (which was the example of Figure~\ref{fig:example-coloring}) is probably the most classic such problem.
Other problems are finding a maximal independent sets, a maximal matching or an edge coloring. 
This is a very active area of research (see \cite{BarenboimE13, HirvonenS20} for recent books on the topic).
As the name suggests, local certification has a strong locality flavor: the nodes have very limited view of the network.
Nevertheless, the concepts originates from another area of distributed computing, self-stabilization.

\paragraph*{Self-stabilization.}
Unlike the classic study of locality, self-stabilization is concerned with faults. It provides an answer to the fundamental question: how to be fault-tolerant? 
The goal is to design algorithms that are self-stabilizing, which means that, starting from an arbitrary configuration, they can converge to a correct configuration~\cite{Dijkstra82}. 
This ensures fault-tolerance because after an arbitrary number of faults, the algorithms will eventually reach a correct configuration.   
A classic book on the topic is \cite{Dolev2000}.
As said above, in self-stabilization, the main challenge is to converge to a correct configuration from an arbitrary one. 
But there is another issue: once you have reached a correct configuration, you would like to stay there. 
In some self-stabilizing contexts, one can move between correct configurations, and the algorithms continues to compute, but being able to stay in the same correct configuration is sometimes a desirable property. 
In \emph{silent self-stabilization}, algorithms should reach a correct configuration, and stay in this configuration, unless a fault occurs. 
In order to be able to detect a possible fault, and also to be sure that the algorithm has stabilized, a very restricted checking procedure continues to be run: every node periodically checks its neighbors to be sure that everything is correct.
We want that if a fault occurs and makes the configuration incorrect, at least one node should be able to detect this while performing its verification. 
This is enough, because this node can then launch a recovery procedure (\emph{e.g.} a global reset), and start the computation of a new correct configuration. 
On the other hand, if the configuration is correct, we do not want to restart computing. 
This way of looking at self-stabilization originates from the seminal works of \cite{AfekKY90, AfekKY97, AwerbuchPV91, AwerbuchV91, AwerbuchPVD94,DolevGS99}.
One can see that the ``stabilized phase'' of a silent self-stabilizing algorithm is basically the scenario described in the introduction.
In the context of self-stabilization, the certificates are pieces of information that are kept in memory during the computation of the solution, to allow the later verification of correctness.  
The example of the path is a bit unusual for this because, for simplicity, we decided to consider a property of the network, and not the correctness of a solution. 
But one can instead consider the classic task of building a spanning tree of the graph. 
Then in addition to pointers to their parents in the tree the nodes will remember their distance to the root, and this will allow them to check that the set of pointers is acyclic (because distances allow detecting cycles). 

\paragraph*{Proof-labeling schemes and locally checkable proofs.}
The notion of proof-labeling schemes has been introduced by \cite{KormanKP10} with the objective of studying the verification phase in details, independently of the self-stabilizing context.  
We will give precise definitions in the next section, but for now, let us say that a proof-labeling scheme is a certification scheme like the one presented in the introduction, described as a pair of algorithms: a (possibly global) algorithm that assigns the certificates and a local algorithm that can verify this certificate assignment.
The paper \cite{KormanKP10} introduced many techniques and problems, and basically started the field of local certification. 
(The certification of acyclicity in the introduction is described in \cite{KormanKP10}, but originates from~\cite{AfekKY97}.)
An important result that was proved in \cite{KormanK07} is that certifying a minimum spanning tree takes certificates of $\Theta(\log^2\!n)$ bits. 
Later, \cite{GoosS16} introduced a generalization of proof-labeling schemes, under the name of \emph{locally checkable proof}. 
The main differences are that the verification can be done at constant distance (instead of distance exactly 1) and that a node can access the identifiers of its neighbors (again, we will give formal definitions in the next section). 
The most important results of \cite{GoosS16} are lower bound techniques adapted to this more general framework. 
An interesting topic, given the historical origin of certification, is to compare the space (in terms of number of bits) needed for certification with the space needed for a full \emph{silent} self-stabilizing algorithm, which would basically build and certify a solution. 
The former is naturally a lower bound for the latter, but in~\cite{BlinFP14}, the authors prove that the two quantities are actually asymptotically equal. That is, one can design self-stabilizing algorithms that do not use more space than the space needed for certification. 
Unfortunately, the construction of~\cite{BlinFP14} requires exponential stabilization time. 

\paragraph*{Distributed decision.}

Around 2012, the idea of building a complexity theory for distributed decision problems was introduced in~\cite{FraigniaudKP13}.
As local certification can be seen as a kind of analogue of non-determinism for decision tasks, it naturally defines an analogue of the complexity class NP. 
Several papers built on this idea, introducing analogue classes for randomized decision, interactive proofs, hierarchies etc. 
These classes actually have several variants depending on the parameters of the model (\emph{e.g.} identifiers and certificate size). 
Section~\ref{sec:complexity} is entirely devoted to this approach.
\vspace{0.5cm}

A lot of results have been obtained after the seminal papers cited in this section. 
We will review some of them in the following sections, and in Section~\ref{sec:directions}  will discuss the most recent works. 
\section{Definitions and vocabulary}
\label{sec:definitions}
In this section, we give more formal definitions for the concepts mentioned so far.

\paragraph*{Graph notions, identifiers, languages, and decision mechanism.}
A graph $G=(V,E) $ is defined by a set of vertices $V$, and a set of edges $E \subseteq V \times V$. 
For distributed graph algorithms, the classic setting is to consider that the network is represented by a  connected graph (that is, for every pair of vertices $u,v$, there is a path in the graph from $u$ to $v$), with no self-loops (that is, no edge of the form $(u,u)$).
The number of nodes is denoted by $n$.
This graph can come with node and edge inputs. 
A graph, possibly with inputs and outputs, is called a \emph{configuration}.
The neighborhood at distance $T$ of a node $v$ is the subgraph induced by all the nodes at distance at most $T$ from $v$.

In the remaining, we will always assume that the nodes have distinct \emph{identifiers} (or ID for short) on $O(\log n)$ bits. 
A graph with identifiers (and possibly inputs and outputs) is an \emph{instance}.
(Some papers in the area deal with networks without identifiers, or consider other identifiers model, but we focus on the standard model.)

We have said that the goal of certification is to allow to check that the configuration satisfies some property. 
One might also use the word \emph{predicate} to refer to a more formal description of a correct configuration.
In general, we follow the literature on centralized computing and consider a \emph{language}: the set of all the correct configurations.
Note that we use the word ``configuration'' here and not ``instance'' ; this is because the languages we consider should not refer to the identifiers. 
For example, a set of graphs where the node with identifier~1 has some special property will not be considered as a language in this paper.
Now, given an instance (that is a graph with identifiers), we will say that it is a \emph{yes}-instance, if when we erase the identifiers, the configuration is in the language considered. Otherwise, the instance is a \emph{no}-instance.

For distributed decision problems, we want to \emph{accept} the \emph{yes}-instances, and to \emph{reject} the \emph{no}-instances. 
Every node is taking its own local decision, \emph{accept} or \emph{reject}, and these decisions are gathered into one global decision. 
Namely, the instance is (globally) accepted if and only if all nodes (locally) accept. 
We will refer to this mechanism as the \emph{decision mechanism}.

\paragraph*{Self-stabilization in the state model.}

We have mentioned that the historical origin of local certification is in self-stabilization. 
Actually, there are several models of self-stabilization, and the one that we were implicitly referring to is the \emph{state model} \cite{Dolev2000}.

\begin{definition}[State model of self-stabilization (partial definition)]
In the state model, every node of the network has a register. Every node has read/write access to its own register, and read access to the registers of its neighbors. 
In one atomic step, a node can read its register and the ones of its neighbors, perform computation, and update its register. 
The register contains all the variables used by the algorithm, including the program counter. 
It neither contains the identifier of the node, nor the program itself.
The register has a special variable which eventually contains the output of the computation.

In this model, an algorithm for some task is self-stabilizing if, starting from an arbitrary collection of registers, the system reaches the set of correct configurations (that is, configurations where the output variables form a solution of the task), and stays in this set.
\end{definition}

Note that the definition above is not complete.
For simplicity, we have only detailed the parts of the state model that are relevant for local certification.
In particular, we have not mentioned the precise model of identifiers (which is the same as for certification), have not defined convergence, and have not stated when the nodes can apply a rule (which are two rather subtle issues that are not essential for certification). 
We refer to \cite{Dolev2000} for a full definition.

\paragraph*{Models of local certification.}

There exists several precise models of local certification. 
The following definition is a common framework for these different models.  

\begin{definition}
A local certification of a given language is described by a distributed algorithm that has the following behavior:
\begin{itemize}
\item On a yes-instance, there exists a labeling of the nodes such that all nodes accept.
\item On a no-instance, for any labeling of the nodes, at least one node rejects.
\end{itemize} 
\end{definition}

The labels are usually called \emph{certificates} or \emph{(local) proofs}. 
The different forms of local certification correspond to different requirements on the algorithm. We now list the three main such forms.

\begin{definition}[Proof-labeling scheme \cite{KormanKP10}]
A proof-labeling scheme is a local certification where the algorithm has access to its identifier, its input (if there is one), its certificates, and the certificates of its neighbors.
\end{definition}

\begin{definition}[Locally checkable proofs \cite{GoosS16}]
A locally checkable proof is a local certification where the node has access to all the information available in its neighborhood at distance $T$, for some constant $T$: all the identifiers, inputs, certificates, and the structure of the graph.
\end{definition}

\begin{definition}[Non-deterministic local decision \cite{FraigniaudKP13}]
A non-deterministic local decision scheme is a proof-labeling scheme where the certificates should not depend on the identifier assignment.
\end{definition}

Note that the literature is not always consistent with respect to the use of these three terms. Also, the definition given here are not written in the same way as in the original papers.

A useful scenario to reason about certification is with a prover and a verifier. 
The prover assigns the certificates. 
It has unlimited power, and wants the nodes to accept, independently of whether the instance is a \emph{yes}-instance or a \emph{no}-instance. 
The distributed algorithm is a distributed verifier that differentiates between the \emph{yes}-instances where the prover is helping, and the \emph{no}-instances where the prover is trying to fool the nodes.

\begin{definition}[Certificate size]
The certificate size of a given labeling of a graph is the size of the largest label. Given a local certification, the certificate size on a given \emph{yes}-instance is the smallest certificate size of a labeling that makes all nodes accept.
Now the certificate size of a local certification is a function $f$ of $n$, such that for every graph size $n$, $f(n)$ is the maximum over all the yes-instances of size $n$ for the certificate size for this instance.
\end{definition}

We will say that a language has optimal certificate size $\Theta(f(n))$ if there exists a local certification with certificate size $O(f(n))$ and there is no local certification with certificate size $o(f(n))$. 
Note that we have not been precise about which model of local certification we consider here. 
This is because for most languages, we can get the best of both worlds: an upper bound in the most constrained model (proof-labeling schemes) and a matching lower bound in the most general model (locally checkable proofs). 
(Non-deterministic local decision has a strictly weaker power in the sense that it can decide a set of languages that is strictly smaller, and it is usually not considered when studying specific languages.)

A concept close to local certification is the notion of \emph{advice}. 
In a distributed environment, an advice also comes as an assignment of labels to the nodes, but these labels can be trusted. 
That is, the information just helps for the computation, and does not need to be verified. 
This concept has been introduced to measure how much global information is needed to complete some task. 
See, for example, \cite{FraigniaudIP06, FraigniaudGIP09,
FraigniaudIP08}.
\section{Certification size: bounds and techniques}
\label{sec:size}

In this section, we review the bounds on the certification size, and the techniques used establish these bounds.
We first show that every language can be certified. 
The challenge is then to establish what is the optimal certification size for interesting languages. 
We describe the classic upper bound techniques in Section~\ref{subsec:upper-bound} and lower bound techniques in Section~\ref{subsec:lower-bound}.
We also provide a table of some important results about certification size in Section~\ref{subsec:landscape}.

\subsection{Universal certification}
It is known that every language can be certified~\cite{KormanKP05}. 
The technique to prove this statement is sometimes called the \emph{universal certification}, and was first mentioned in~\cite{KormanKP05}. 
On \emph{yes}-instances, the prover provides the whole adjacency matrix of the graph, with the correspondence between the rows and the identifiers of the nodes (and also the input assignment if there are inputs). 
The nodes first check that they are given the same certificate. 
Then they check that this description of the graph is consistent with their local views.
It is easy to show that if this step succeeds, then the instance described in the certificates is indeed the real instance. 
Finally, all the nodes check, in parallel and without communication, that the configuration is in the language.  
This scheme uses certificates of size $\Theta(n^2)$. (If there are inputs on the nodes or edges, then these also have to be encoded and take additional space. 
As these are typically of constant size, the size is still in $\Theta(n^2)$.)
\subsection{Upper bounds techniques}
\label{subsec:upper-bound}
The techniques used in certification depend of course on the language being certified. Nevertheless, we can identify a few techniques that are very common.
\paragraph*{Basic techniques.}
The definition of some languages basically describes their certification. 
For example, to certify that a graph can be properly $k$-colored, one just has to label the nodes with their colors, as the nodes can check locally that the coloring is correct. 
More generally, for any LCL problem (as defined in Section~\ref{sec:history}) the set of graphs that admit a solution can be certified by simply writing the solution in the certificates.
Another property that is easy to certify is the fact that some node has some special role. 
More concretely, consider that the nodes are given a bit as input, and that we want to certify that at most one node is given 1. 
One can basically give the identifier of the selected node to all nodes. More precisely, consider the following local decision algorithm on a node $v$: 
\begin{enumerate}
\item If the certificate of a neighbor of $v$ is different from the certificate of $v$, then reject.
\item If $v$ has input 1, and the certificate is not the identifier of $v$, then reject.
\item Otherwise, accept. 
\end{enumerate}
On a \emph{yes}-instance, either there is one node with a 1, and the prover labels every node with the identifier of this node, or there is no such node, and the prover can give empty labels. 
In both cases, all nodes accept. 
On a \emph{no}-instance, there are at least two nodes with a 1. 
Then there are two situations. 
First, one of them does not see its identifier in the certificate, and then it rejects. 
Second, they both see their own identifier in their certificate. 
Then we can partition the graph into several non-empty regions where the nodes have the same certificate, and by connectivity, there exists an edge that has two endpoints that do not have the same certificate. (Remember that the identifiers are all distinct.) These nodes reject because they do not have the same certificate. 
\paragraph*{Spanning trees.}
A central tool in certification is the one of spanning trees.
Consider that the graph is given as input a binary labeling of the edges. 
We want to check that the selected edges (the ones labeled with 1) form a spanning tree of the graph. 
On \emph{yes}-instances, the prover can assign the certificates in the following way \cite{AfekKY90}. 
\begin{enumerate}
\item Chose a node to be the root of the tree. 
\item Write the identifier of the root in all certificates.
\item Give to every node its distance to the root in the spanning tree.
\end{enumerate}
The local decision algorithm is the following.
\begin{enumerate}
\item Check that the root identifier is the same that the one given to the neighbors. 
\item If the distance is 0, check that the root identifier of the certificate is the same as the identifier of the node.
\item If the distance is 0, also check that all neighbors linked by a selected edge have distance 1.
\item If the distance $d$ is not 0, check that \emph{among the neighbors to which the node is linked by selected edges}, there is exactly one node with distance $d-1$, and the other such neighbors have distance $d+1$. 
\end{enumerate}
It is clear that on a \emph{yes}-instance, all nodes accept. 
For a \emph{no}-instance, consider first the case where the set of selected edges do not form an acyclic subgraph. 
In this case, consider a cycle $C$ of selected edges. In $C$ consider a node $v$ with the largest distance. 
This node has two neighbors in $C$ and none of them has a strictly larger distance. 
Then $v$ rejects, either in step 3 if it has distance $0$, or in step 4 if it has distance strictly larger.    
Now consider the case where we have a \emph{no}-instance with an acyclic but disconnected set of selected edges. 
In every tree, at least one node must have distance 0, otherwise one node would reject at step 4. And if several nodes have distance 0, we are back to the case we described above: thanks to the root identifier given to every node, at least one node should reject.
This local certification uses certificates of size $O(\log n)$ because the distances are between 1 and~$n$, and the identifiers are assumed to be encoded on $O(\log n)$ bits.  
Spanning trees are everywhere in local certification because they allow to certify that there is a unique node having some property, which is very useful. 
To certify this, on \emph{yes}-instances, the prover chooses a spanning tree rooted at the special node, gives to every node the identifier of its parent in the tree, and certifies the spanning tree as explained above.   
Then the root checks that indeed it has the property, and all the other nodes check that they do not have this property.
A spanning tree also allows some counting in the graph \cite{KormanKP10}. For example, suppose once again that the nodes have a bit as input, and that we want to check that there are $k$ 1s, for some $k$ specified in the language. 
The prover can describe and certify a spanning tree, and give to every node $v$ a counter, which contains the number of 1s there is in its subtree of the spanning tree rooted at $v$. 
Then the nodes can check the consistency of the counters: the counter of a node should be the sum of the counters of its children, plus its own input bit. 
The root can check that its counter is $k$. 
Using this tool, we can, for example, certify the number $n$ of nodes in the graph, or the size of a structure that can be checked locally, such as a clique, an independent set, or a matching.
 
\begin{figure}[!h]
\begin{center}
\begin{tabular}{ccc}
Round 0 & Round 1 & Round 2 
\\
\\
\includegraphics[scale=0.7]{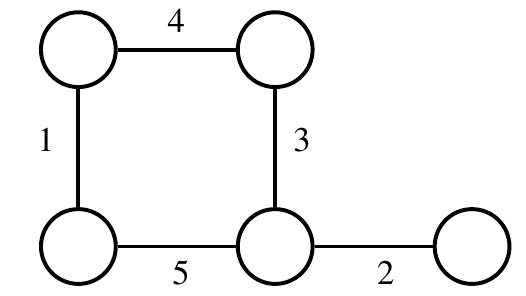}
&
\includegraphics[scale=0.7]{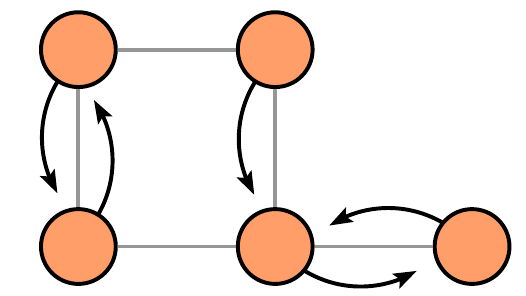}
&
\includegraphics[scale=0.7]{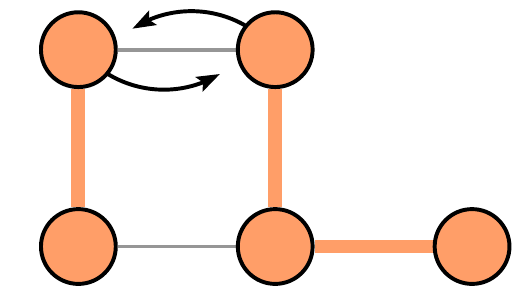}
\\
\\
\includegraphics[scale=0.7]{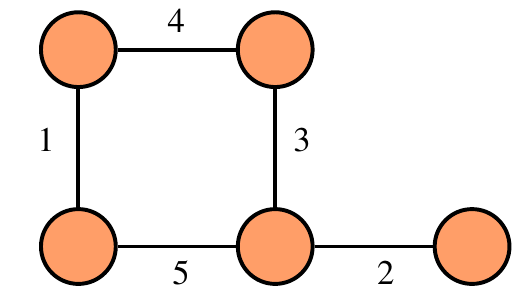}
&
\includegraphics[scale=0.7]{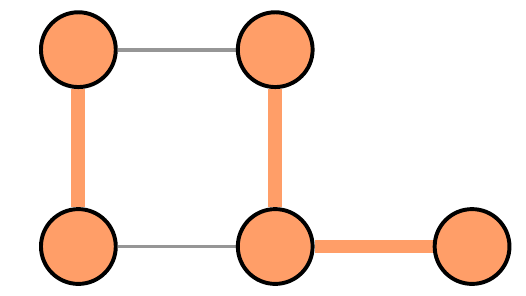}
&
\includegraphics[scale=0.7]{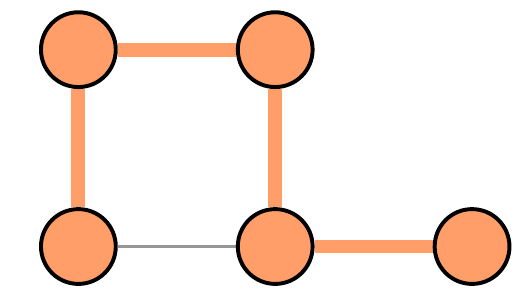}
\end{tabular}
\end{center}
\caption{\label{fig:boruvka} 
Several minimum spanning tree (MST) algorithm are based on merges of so-called fragments, from the historical Borůvka algorithm \cite{NesetrilMN01} to the celebrated GHS algorithm \cite{GallagerHS83}.
The run of such an algorithm is described in this figure. 
Very roughly, every node starts as its fragment, and at each round, each fragment proposes to merge with the neighboring fragment to which it has the lightest edge. 
At the end of the process, these merging edges form a minimum spanning tree.
As the number of fragments is at least halved at each step, there are at most $O(\log n)$ rounds. 
In other words, a node belongs to at most $O(\log n)$ successive fragments. 
The certificate at some node~$v$ consists in $O(\log n)$ fields, one for each fragment the node belongs to. 
For each field, the node is given a pointer to a neighbor (in the MST), a distance and an identifier, such that collectively the certificates of the nodes of a fragment form a spanning tree. This spanning tree is pointing to the node through which the next merge will be performed.
Also, every node is given the weight and name of the successive merge edge of its fragments. 
This is enough for certification: the nodes can check the fragment structure and also the fact that the merges happen on the lightest outgoing edges. 
}
\end{figure}

\paragraph*{Encoding the run of an algorithm.}
A technique that already appears in the original proof-labeling scheme paper \cite{KormanKP10} consists in encoding the run of an algorithm.
Consider for example the problem of certifying that a set of edges forms a minimum spanning tree, and suppose you have a distributed algorithm that computes this solution.  
Then a possible certification consists in encoding at every node, every step of the algorithm. 
For example, at round 1, the algorithm sent this message to that neighbor and received that other message from that other neighbor, and then at the second round this and that happened etc. 
Given the code of the distributed algorithm, the node can always check that the run is correct and consistent (the messages that are said to be sent by a node are written as received by the other nodes etc.). 
If every node is happy with the run, and if the run indeed selects the edges of the input, then the input describes a minimum spanning tree. 
In general, this technique is not very efficient in terms of certificate size. 
This is because a lot of information exchanged during the algorithm is redundant or just useless for the verification of the solution.  
Nevertheless, this approach can be fruitful.  
For example, one can certify some data structures that are used to build the solution, without explicitly writing all the messages sent to build this structure.
For the problem of minimum spanning tree, this technique is actually optimal (for the usual weight range) \cite{KormanKP10}. See Figure~\ref{fig:boruvka} for an overview of this certification. 
\paragraph*{Duality.}
When it comes to certifying some combinatorial structures, one can sometimes use duality. 
Let us take the example of maximum matching in bipartite graphs \cite{GoosS16}.
König's theorem states that on bipartite graphs, the size of the maximum matching is equal to the minimum size of a vertex cover. 
We can be more precise and say that a matching is maximum if and only if there exists a vertex cover that consists in one endpoint of each edge of the matching.
Then, to certify that a set of selected edges is a maximum matching, the prover can simply put a bit on each vertex saying whether it is in a vertex cover. The nodes can then check locally the vertex cover, the matching, and the fact that every edge of the matching has one node of the vertex cover. And this is enough.
On this example, the duality was implicit. To certify a maximum weight matching in a weighted bipartite graph, one explicitly writes the values of  the primal and dual variables \cite{GoosS16}.
Actually similar but more involved techniques allow one to certify maximum cardinality matchings and 2-approximation of maximum weight matching in general graphs \cite{Censor-HillelPP20}. 
A lot of other approximation problems have been certified using this technique \cite{EmekG20} (we will come back to this in the last section).

\paragraph*{Other techniques.}
Other techniques have been used for specific problems, that could possibly be used in other problems. 
For example, the certification of minimum spanning tree in $O(\log n \log W)$ bits for weights in $[1,W]$ from \cite{KormanK07} uses a variety of ideas. 
For example, when using some $k$ spanning trees to point to $k$ different nodes, instead of using $k \times \Theta(\log n)$ bits, one can use one spanning tree on $O(\log n)$ bits, and then use it to encode the other spanning trees using only $O(1)$ bits per additional tree. See \cite{Feuilloley19-note} for an explanation of the other tricks of \cite{KormanK07}. 
In general, techniques from other labeling problems (distance labeling, adjacency labeling) can also be useful in certification.
 
\subsection{Lower bound techniques}
\label{subsec:lower-bound}
\begin{figure}[!h]
\begin{center}
\includegraphics[scale=1.5]{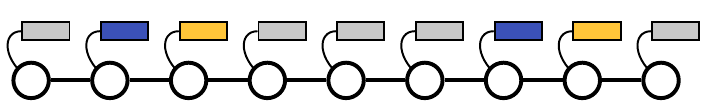}\\
\includegraphics[scale=1.5]{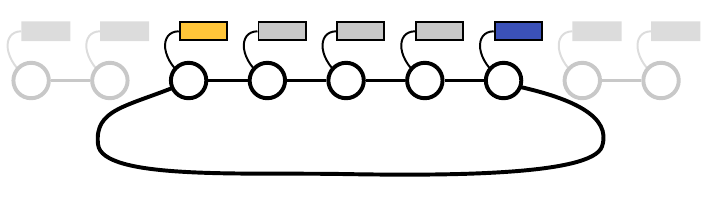}
\vspace{-0.7cm}
\end{center}
\caption{\label{fig:crossing-technique} 
This figure illustrates the \emph{crossing technique} for the language of paths, in the proof-labeling scheme model~\cite{KormanKP10}. 
The idea is that if $o(\log n)$-bit certificates are used, then by the pigeon-hole principle (or equivalently by counting argument), there exist two edges, $(a,b)$ and $(c,d)$, such that $a$ and $c$, respectively $b$ and $d$, are assigned the same certificate. 
These shared certificates are colored blue and yellow in the top picture.
Then we \emph{cross} the edges: we remove $(a,b)$ and $(c,d)$, and add $(b,c)$. 
We get a cycle, which is a \emph{no}-instance, and some disconnected parts that can be discarded. 
The point is that for all nodes in the cycle, even the ones that are adjacent to the crossed edge, the view is the same as in the path, thus all nodes accept. And this is a contradiction.
}
\end{figure}

\begin{figure}[!h]
\begin{center}
\begin{tabular}{cc}
\includegraphics[scale=0.68]{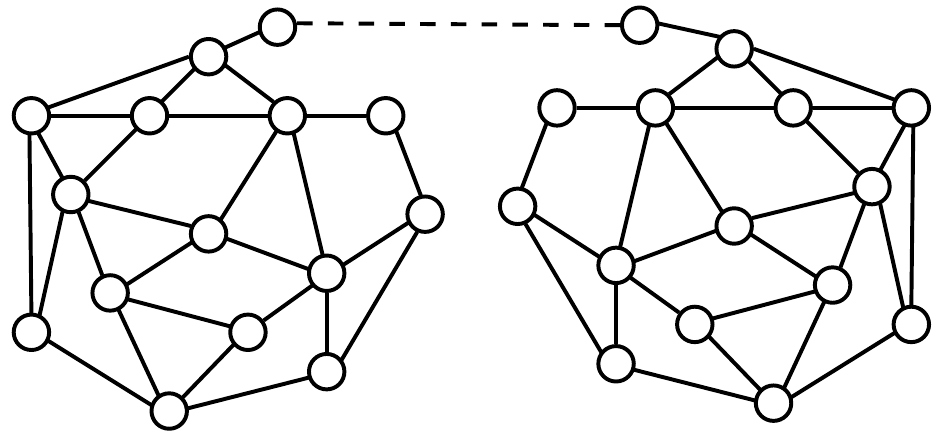}
&
\includegraphics[scale=0.68]{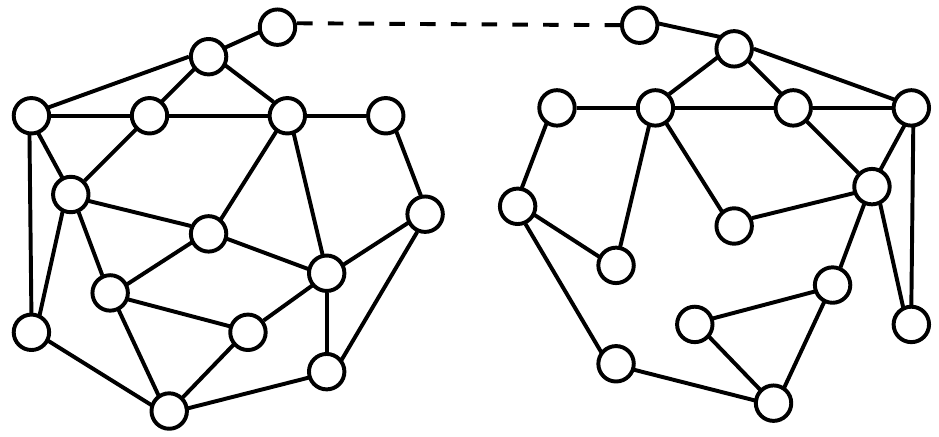}
\\
(a) Symmetric dumbbell graph
&
(b) Non-symmetric dumbbell graph
\end{tabular}
\end{center}
\caption{\label{fig:symmetric} Illustration of the symmetric dumbbell graph language.}
\end{figure}

There are basically two types of techniques used for lower bounds: modifications of one or several instances to get to a contradiction via a counting argument, and reductions from communication complexity (that can be explicit or implicit). 
The first one works for $\Omega(\log n)$ bounds, and the second is more adapted for higher lower bounds.
An exception is \cite{KormanK07}, where the authors use techniques tailored for the minimum spanning tree problem.
\paragraph*{Crossing technique.}
In terms of lower bounds, there is a difference between proof-labeling schemes and locally checkable proofs. 
It is easier to derive lower bounds in the first model because the nodes are less powerful, thus can be fooled more easily. 
A technique for lower bounds in proof-labeling schemes is the crossing technique.
In this technique, we take a \emph{yes}-instance and show that we can modify it in such a way that (1) the nodes do not detect a change (and then continue to accept) and (2) the instance is now a \emph{no}-instance~\cite{KormanKP10}. 
See Figure~\ref{fig:crossing-technique} for an example. 
Unfortunately, this technique does not work in the more general model of locally checkable proofs. 
Indeed, in this later model, the nodes can see the identifier of their neighbors, then any modification of one instance implies that the node have different views, and can take different decisions.
\paragraph*{Cut-and-plug technique.}
A type of technique that provides similar bounds to the crossing technique, but for locally checkable proofs, is what we call \emph{cut-and-plug}.
It consists in taking several \emph{yes}-instances, cutting them into pieces, and proving that we can create a \emph{no}-instance by plugging several of these pieces together, in such a way that the nodes cannot detect it.
Let us describe a general framework for this technique for the language of paths. (There exist several implementations of this general framework \cite{GoosS16,FeuilloleyH18}.)
Suppose that there exists a certification with $f(n)$-bit certificates with $f(n)\in o(\log n)$, where the nodes can see at some constant distance $k$ (remember that the locally checkable proofs model allows this). 
Now consider $\Theta(n)$ paths $P_i$ on $2k+1$ nodes, such that any identifier appears in at most one path. 
We will call these \emph{chunks}, and consider that each chunk has an arbitrary orientation, in order to talk about the first node and the last node.
For any given chunk, we can consider a labeled version of it, where every node $v$ is assigned some label of size $f(n)$. 
For all $P_i$, we define $(P_{i,j})_j$ as the set of all the possible such labeled versions of it.  
Now consider two labeled chunks $P_{i,j}$ and $P_{i',j'}$. 
We say that we can plug them, and write $P_{i,j} \rightarrow P_{i',j'}$, 
if when we add an edge between the last node of $P_{i,j}$ and the first node of $P_{i',j'}$, and run the verification, all nodes at distance at most $k+1$ from the added edge accept (the decisions of the other nodes is not important here). 
Consider now a path made by concatenating some chunks $P_i$.
As this is a \emph{yes}-instance, there exists for each chunk $P_{i}$ of the instance, a labeled version $P_{i,j}$, such that all nodes accept. 
Moreover, as the verifier algorithm sees only at distance $k$, for all consecutive labeled chunks $P_{p,q}$ and $P_{p,q'}$ we have $P_{p,q}\rightarrow P_{p',q'}$. 
By considering all the \emph{yes}-instances, one can get a large collection of labeled chunks that can be plugged. 
One can then show via a counting argument that there exists $\ell>2$, and a family $(P_{p^i,q^i})_i$, $i\in[1,\ell]$ such that: (1) for every $i<\ell$, $P_{p^i,q^i} \rightarrow P_{p^{i+1},q^{i+1}}$, and (2)~$P_{p^{\ell},q^{\ell}} \rightarrow P_{p^{1},q^{1}}$.
That is, there is a collection of chunks that we can plug into a cycle, and a collection of labels such that every node accepts. 
This is a contradiction. 
This technique can be implemented in different ways, depending on the definition of the chunks, and on the counting argument. See \cite{GoosS16} and \cite{FeuilloleyH18}.
\paragraph*{Proofs inspired by communication complexity.}
A classic lower bound technique in distributed computing is to design a reduction from communication complexity. 
In certification, a first intuition for why this can work is the following. 
The nodes have to take a decision concerning a global property of the network, and basically the only kind of communication they have is through the certificates. 
Then intuitively, if to decide whether the instance is correct or not, a lot of information has to be transferred from one side of the graph to the other, the certificates have to be large.
For example, Figure~\ref{fig:symmetric} illustrates the language of symmetric dumbbell graphs \cite{GoosS16, KormanKP10}. 
These are graphs formed by taking two copies of the same graph and linking them by a long path (that is, a path of $2k+1$ nodes for verification radius $k$). 
To certify that the graph is in this family, one basically has to write the map of the extremity graphs in the certificate. 
This can be proved by a counting argument. 
Intuitively, the nodes of the paths cannot do clever things, and we can consider that they have the same certificate. 
If this certificate could work for several extremity graphs, then we could take two pairs and mix them into a \emph{no}-instance that would be accepted, and this would be a contradiction. Hence, there is basically one ``path certificate'' per extremity graph, which means the certificates have size $\Theta(n^2)$.  
\paragraph*{Reduction from non-deterministic communication complexity.}
A more advanced lower bound technique uses explicit reduction from communication complexity. 
The reduction is from the classic disjointness problem where each player is given a set, $A$ for Alice, and $B$ for Bob, and the goal is to minimize the number of bits exchanged in order for Alice and Bob to decide whether $A\cap B$ is empty. 
It is well-known that this problem has high communication complexity even in the non-deterministic setting, where the communication can also take the form of labels given to Alice and Bob by a prover \cite{KushilevitzN97}. 
The idea is then to prove that if a local certification with small certificates existed for the problem considered, then one could design a (non-deterministic) communication protocol for disjointness that would beat the known lower bounds.
Basically, in this communication protocol, Bob and Alice would simulate the verification process on a specific graph depending on their inputs.
Bob would be in charge of simulating half of the nodes, and Alice would be in charge of simulating the other half, and the communication would be used to exchange the certificates of the nodes on the cut between Alice's and Bob's vertices. 
The exact construction depends on the language studied, because the graph should encode the disjointness problem with gadgets depending on that language. 
To our knowledge, there are two examples of such proofs in the literature: an $\Omega(n^2/\log n)$ bound for non-3-colorability~\cite{GoosS16} and an $\Omega(n/k)$ bound for diameter at most $k$ in \cite{Censor-HillelPP20} (see also \cite{FeuilloleyFHPP20}). (For an introduction to this type of techniques, see \cite{Feuilloley21}.) 
\subsection{Landscape of proof sizes}
\label{subsec:landscape}
To finish this section, we give a small table of some important bounds on certification size. See Table~\ref{table:sizes}. 
\begin{table}[!h]
\begin{center}
\begin{tabular}{c|c|c|c}
Language 
& Upper bound 
& Lower bound
& Reference\\
\hline
Locally checkable language
& 0
& --
& \cite{NaorS95} \\
$k$-colorability
& $O(\log k)$
& --
& Folklore \\
Maximum matching in general graphs
& $O(\log n)$
& $\Omega(\log n)$
& \cite{Censor-HillelPP20, GoosS16} \\
Spanning tree
& $O(\log n)$
& $\Omega(\log n)$
& \cite{KormanKP10, GoosS16} \\
Minimum spanning tree (weights in $[1,W]$)
& $O(\log n \log W)$
& $\Omega(\log n \log W)$
& \cite{KormanK07}\\
Diameter $k$
& $O(n\log n)$
& $\Omega(n/k)$
& \cite{Censor-HillelPP20} \\
Non-3-colorable graph
& $O(n^2)$
& $\Omega(n^2/\log n)$
& \cite{GoosS16}\\
Symmetric graph
& $O(n^2)$
& $\Omega(n^2)$
& \cite{GoosS16, KormanKP10}\\
Any language
& $O(n^2)$
& --
& \cite{KormanKP10}
\end{tabular}
\end{center}
\caption{\label{table:sizes}Some important certification bounds. All upper bounds hold in the proof-labeling scheme model. 
All lower bounds hold in the locally checkable proofs model, except the ones for minimum spanning tree and  diameter that hold in the proof-labeling scheme model.}
\end{table}

Note that the languages that require no certificate or very small ones, such as different versions of coloring, correspond to properties that are usually considered as local. 
On the other hand, languages that require large certificates, such as symmetric graphs, correspond to properties that are considered global. 
Therefore, at least on an intuitive level, the certificate size is a measure of locality: the smaller the certificates, the more local the property.

\section{Distributed decision: a complexity theory point of view}
\label{sec:complexity}
As mentioned in earlier sections, one way to look at certification is to consider that it is the non-deterministic version of distributed decision. 
This analogy appears already in \cite{KormanKP10, GoosS16}. 
In \cite{FraigniaudKP13} the authors pushed the idea further by  formally introducing several new classes, and proving inclusions between these classes. 
(To be precise, \cite{FraigniaudKP13} is also different because it is about distributed decision without reference to identifiers.) 
As in the current paper we focused on certification, we will not give an exhaustive survey of this area, but it is nevertheless insightful to have a quick overview. 
We refer to \cite{FeuilloleyF16} for a survey on distributed decision.
\paragraph*{Deterministic and probabilistic distributed decision.}
As mentioned in the introduction, there are languages that can be recognized locally. 
One example is the set of graphs properly colored with $k$ colors. These languages form the class of local decision LD \cite{FraigniaudKP13}, which is basically the same as the classic LCL \cite{NaorS95}. 
Many papers are interested in constructing solutions for such languages (problems such as $(\Delta+1)$-coloring, maximal matching, independent set). 
Actually, there is now a complexity theory for the construction problems whose solutions are LCL, see for example \cite{Suomela20}. 
Note that this is not the same as what we are describing now: we are currently interested in decision problem, not construction. 
Unlike in centralized computing, as far as we know, the two domains have little in common.
Deterministic decision (that is, local decision without certificates) only captures a rather small set of languages, and \cite{FraigniaudKP13} introduced the idea of looking at a probabilistic version.
It consists in allowing nodes to use random bits when taking decision, with the goal that on \emph{yes}-instances, there is a good probability of acceptance, and on \emph{no}-instances, there is a good probability of rejection. 
Note that in probabilistic decision, like in deterministic decision, there are no certificates.  
The classic example for probabilistic decision in the language where nodes are given binary labels, and at most one node should have a 1 \cite{FraigniaudKP13}. 
Clearly, this cannot be decided with a deterministic decision. 
Here is a distributed probabilistic decision algorithm for this problem.
Every node with a 0 accepts, and every node with a 1 accepts with probability $p=(1+\sqrt{5})/2$. 
On instances where there is no 1, this algorithm is always correct. On instances where there is one 1, it is correct with probability $p$. 
On instances where there are more than one 1, the algorithm should reject, and this happens with probability at least $1-p^2$, which for our choice of $p$ is also $(1+\sqrt{5})/2$. 
That is, this decision algorithm is correct with probability $(1+\sqrt{5})/2$.
One should note that unlike what happens in centralized computing, the precise thresholds is important, as boosting is not always possible~\cite{FraigniaudGKPP14}. 
Indeed, in centralized computing, the classic boosting technique consists in running the probabilistic decision several times, and taking the most frequent outcome, which it is not possible in the distributed local setting, as the nodes do not know the outcome of the decision process. 
 
Probabilistic decision is useful even outside the complexity perspective, as it gives a natural candidate when one wants to generalize results from LCL to a larger class. 
For example, a seminal result in the LOCAL model is that constant-time randomized algorithms for LCL can always be derandomized~\cite{NaorS95}, and this result has been generalized to languages that can be decided via a probabilistic decision~\cite{FeuilloleyF15}.
\paragraph*{Non-deterministic distributed decision.}
As said, certification is a form of distributed non-determinism. 
Remember that the complexity class NP has two classic definitions: the one with a non-deterministic Turing machine, and the one where polynomial size certificates are assigned by a prover \cite{AroraB09}. 
The second definition is clearly the one for which the analogy is the most natural.
Note that for the class NP, both the size of the certificates and the computational power are limited. 
For now, we have only mentioned restrictions on the analogue of the computational power, which in our model is the size of the view of a node, and we have set this to constant.
As for the certificate sizes, instead of limiting it, we have measured how large they need to be for various languages. 
In \cite{GoosS16}, the authors defined LogLCP as the class of languages that have certificates in $O(\log n)$, and argue that this forms a meaningful class. 
Indeed, there are many languages that fall into this class, and we know very few languages between certificates of size $O(\log n)$ and $\Omega(n)$. (Basically only minimum spanning tree is a natural such languages).
\paragraph*{Hierarchy and interactive proofs.}
As in centralized computing, we can consider more elaborate computational settings than non-deterministic distributed decision. 
In this direction, \cite{FeuilloleyFH21, BalliuDFO18} introduced analogues of the polynomial hierarchy in complexity theory (see \emph{e.g.} \cite{AroraB09}). 
The idea is that in addition to the prover trying to convince the nodes that the instance is a \emph{yes}-instance, there is disprover trying to convince the nodes that the instance is a \emph{no}-instance.
The certificates given by these players are of size $O(\log n)$, and they assign them to the node one after the other. 
For example, on the third level of the hierarchy, the nodes first receive certificates from the prover, then receive certificates from the disprover (that are somehow answers to the prover's certificates), and then again from the prover. The number of alternations defines a hierarchy.
An analogue of Arthur-Merlin interactive proofs has also been proposed in \cite{KolOS18}. 
There the setting is that there is a unique prover, but the nodes have access to randomness, and there can be several rounds of communication between the prover and the nodes. 
This model provides a better understanding of problems such as symmetric dumbbell graphs in Section~\ref{subsec:lower-bound}, and follow-up works have shown that powerful tools can be designed for this setting, with trade-offs between the different resources (number of bits exchanged, number of prover-nodes communication rounds etc.) \cite{CrescenziFP19, NaorPY20, MontealegreRR20}. 
\section{Research directions}
\label{sec:directions}
We now list some research directions, mentioning recent results and some open questions. 
\subsection{New upper and lower bounds}
An evident research objective is to establish the optimal certification size of any interesting language. 
For classic problems of distributed computing, such as tree structures or connectivity questions, tight bounds have been known for a decade or more now, but we are far from a complete understanding of this type of question. 
Let us start with two simple open questions.

\begin{open}
The property of $k$-colorability is easy to certify with $O(\log k)$ bits, just by writing the colors. Can we prove a matching lower bound?
\end{open}

\begin{open}
In \cite{Censor-HillelPP20}, the authors show that the optimal size for certifying diameter at most $k$ is in $\Omega(n/k)$ and $O(n\log n)$. 
What is the optimal certification size?
\end{open}

For a lot of problems, either we have no idea what the certification size is, or we have polylogarithmic certification. 
The only polynomial lower bounds we have are for symmetric graphs, 3-non-colorability \cite{GoosS16}, diameter \cite{Censor-HillelPP20} and for the uniqueness language (where all nodes should have different inputs)~\cite{KormanKP10}. 
Hence, the following problem: 
\begin{open}
What are other interesting languages with polynomial optimal certification size?
\end{open}
A recent topic in certification is the idea of approximate certification. 
This was introduced by~\cite{Censor-HillelPP20}, with the following striking example. With the classic definition, certifying that the diameter is at most $k$ takes $\Omega(n/k)$ bits. 
Now, if we relax the task, and ask that all graphs of diameter at most $k$ are accepted, and all graphs of diameter at least $2k$ are rejected, then the certificate size drops to $O(\log n)$ bits. 
One can consider a lot of problems from the approximation perspective: \cite{Censor-HillelPP20} also studied maximum matching and \cite{EmekG20} made a systematic use of primal-dual methods for various optimization problems. 
The paper \cite{EmekG20} also introduced the idea of limiting the computational power of the prover and the verifier, and deriving conditional lower bounds from hypothesis such as P$\neq$NP.
It is also interesting to think about what can be certified outside the classic graph problems (coloring, spanning trees, matching etc.). 
For example, \cite{BalliuF19} introduced the topic of certifying routing tables. 
It is also very interesting to study the properties of the graph itself, and then one can see the language as a graph class. 
There are many interesting graph classes, and establishing the certificate size for those is a vast program. 
In \cite{FeuilloleyFMRRT21} and \cite{FeuilloleyFMRRT20-b}, the authors proved that certifying planarity and embeddability on bounded genus graphs can be done with certificates of size $O(\log n)$. 
Note that these classes can be characterized by forbidden minors, and  that the language that we know have large certificates (\emph{e.g.} diameter~$k$) do not have such characterizations. 
The following question appears natural.
\begin{open}
Can we certify all graph classes characterized by forbidden minors with certificates of size $O(\log n)$? 
\end{open} 
For the question above, even if the answer is positive, the constant of the asymptotics will probably depend on the size of the excluded minors.
This brings us to the next question: what happens if we express the certification size not only as a function of the size $n$ but also as a function of other parameters?
In other words, what is the parameterized complexity of distributed certification?
There exists some bounds using such parameters, but we could use those much more systematically.
Natural parameters are the diameter, the maximum degree, the arboricity, the number of edges and the treewidth. 
Note that taking this point of view would require new techniques, not only in upper bounds but also in lower bounds, where (depending on the parameters) one would not be able to use the classic ring or small cuts graphs, that are used respectively for cut-and-plug and for reduction from communication complexity. 
Here are two concrete problems. 
\begin{open}
Establish the optimal certification size of the diameter as function of the number of nodes~$n$ and of the diameter $k$. Establish the optimal certification size of bounded-genus graphs as function of $n$ and of the genus $g$. 
\end{open}
Some certification work only in some specific graph class, for example, maximum weight matching in bipartite graph \cite{GoosS16} and diameter on trees \cite{KormanKP10}. 
It is a natural research direction to try to prove such results, and to find where the limit is between compact and polynomial certification.
\begin{open}
For the languages that have polynomial optimal certification, find a large graph class in which this language can be certified with polylogarithmic certificates. 
\end{open}
Finally, even for bounds we know, it would sometimes be nice to have an alternative proof.
\begin{open}
Is there a simpler proof for $\Omega(\log n \log W)$ bound for minimum spanning tree established in \cite{KormanK07}?
\end{open} 

\subsection{Understanding certification}

The study of the local certification is not limited to establishing upper and lower bounds on the certificate size. 
Another way to gain understanding on the notion is to modify the model, for example  by increasing or reducing the resources allowed.

First, let us consider the communication resources.
In standard proof-labeling schemes, the communication model follows the state model of self-stabilization: a node can see the states of its direct neighbors. 
This can be seen as an abstraction of a model where every node sends its states to all its neighbors at every round.
This aspect can be modified in at least two ways. 
First, in~\cite{Patt-ShamirP17}, the authors proposed to consider a message-passing model, with different messages for different neighbors.
Second, in~\cite{FraigniaudPP19}, randomization was used to reduce the size of the messages.  

Still on the communication side, another line of work proposes to increase the view of the nodes beyond constant \cite{OstrovskyPR17, FeuilloleyFHPP20} (see also~\cite{KormanKM15}). 
In other words, these papers study the impact of a larger verification radius on the certificate size.
The following question is still open, although partial positive answers have been obtained.

\begin{open}
Can we always achieve a linear trade-off between the certificate size and the verification radius?
\end{open}

It is also relevant to challenge the certification side of the model. 
In \cite{FeuilloleyH18}, the certification is not local, that is with one certificate per node, but global, that is, all nodes can access the same unique certificate.
A simple open question in this model is:
 
\begin{open}
Show that certifying bipartiteness requires a global certificate of size $\Omega(n\log n)$.
\end{open}

Interestingly, the results about larger verification radius and global certificates shed a new light at the notion of redundancy in certification. 
On the one hand, the fact that increasing the radius allows to decrease the size of the proofs indicates that there is some redundancy in the certificates. 
For example, it is not the case that in total we need $n \times O(\log n)$ bits to certify a spanning tree, because if we allow a view at logarithmic distance, \cite{FeuilloleyFHPP20} proved that we are fine with $O(1)$-bit certificates. 
On the other hand, this redundancy is not global, in the sense that it is not the case that the exact same information is copied in every certificate. 
Indeed, \cite{FeuilloleyH18} proved that a global certificate for spanning tree cannot be smaller than the collection of all the local certificates, that is, must be of size $\Omega(n \times \log n)$.

Yet another direction to modify the model is to restrict the computational power of the nodes.
In this direction, a fruitful approach is to use modal logic to characterize the languages that can be recognized. See for example the thesis~\cite{Reiter17}, and the references therein.
See also \cite{EmekG20} for an example where limiting the computational power to polynomial-time allows deriving better lower bounds.

Finally, we have seen that the decision mechanism ``accept globally if and only if every node accepts locally'' \cite{AfekKY97} is justified by the origins of certification in the context of self-stabilization. 
But it is reasonable to ask about other models where the mechanism would be different.
For example, \cite{ArfaouiFP13} and~\cite{ArfaouiFIM14} investigated contexts where the nodes can output more than one bit, that is, the set of possible outputs is strictly larger than just \emph{accept} or \emph{reject}. 
The global decision is then an arbitrary function of these outputs. 
An other example where the decision mechanism is a bit different is \cite{FeuilloleyF17}, where it is required that in instances that are far from being in the language, many nodes should reject. 

\subsection{Complexity theory}
As described in Section~\ref{sec:complexity}, there has been efforts to build a complexity theory for distributed decision. 
Here is one open question from one of the papers we cited.
\begin{open}
Prove that the decision hierarchy of \cite{FeuilloleyFH21} is infinite.
\end{open}
Note that there are evidence that this question might be a very hard~\cite{FeuilloleyH18}.
We have also mentioned distributed Arthur-Merlin protocols introduced in ~\cite{KolOS18}. This is a recent and active area of research. We refer to \cite{NaorPY20} for a list of nice open questions. 
Let us just mention that an important question is to establish the best trade-off possible between interaction and communication. 
As we gain knowledge about distributed complexity classes, one might want to go one step further and ask to characterize such classes. Here are two such questions.
\begin{open}
Can we characterize the languages that have a probabilistic decision scheme? What about the ones that have $\Theta(\log n)$ certification? 
\end{open}
Finally, we have seen in the definitions that there are different types of local certification: proof-labeling schemes \cite{KormanKP10}, locally checkable proofs \cite{GoosS16} and non-deterministic distributed decision \cite{FraigniaudKP13}.
One of the differences between these models is the way they handle identifiers. 
This difference in turns implies different associated complexity classes. 
For example, \cite{FeuilloleyFH21} and \cite{BalliuDFO18} study hierarchies in two models that differ by their use of the identifiers, and these hierarchies end up being completely different.
The impact of the precise model of identifiers on decision can be a pretty subtle issue, as shown for example in \cite{FraigniaudHS18, FraigniaudHK12, FraigniaudGKS13}.
\subsection{Certification in self-stabilization}
Originally, certification has been studied in the context of self-stabilization, and can be identified as a component of silent self-stabilizing algorithms. 
There are still interesting interactions between the two. 
First, the usefulness of studying  certification to better understand self-stabilization has been illustrated recently, with the celebrated minimum spanning tree certification of \cite{KormanK07} being embedded into a full self-stabilizing algorithm \cite{BlinDF20}. 
A surprising question actually arises from \cite{BlinDF20}. 
In (silent) self-stabilization\footnote{Note that we highlight the relation between silent self-stabilization and certification, because the two are very close, but non-silent self-stabilization also uses some ideas that we discuss, including local decision.}, a typical algorithm would keep some pieces of information during the computation of a solution, and these pieces will serve as a certificate. 
Instead, in \cite{BlinDF20} a solution is first built, and only then certificates are computed. 
There are evidences that this modular approach is necessary \cite{Feuilloley20}, but nothing has been formally proved, and we even lack a precise definition.
\begin{open}
Formalize the notion of \emph{certification during computation}, and show that it cannot be achieved for minimum spanning tree in space $O(\log n \log W)$. 
\end{open}
In \cite{BlinFP14}, the authors proved that one can always design a silent self-stabilizing algorithm that does not use more than the space needed for certification. 
Unfortunately, this general transformation takes exponential convergence time in general. Hence, the following general question.
\begin{open}
In the state model of self-stabilization, when can we use optimal space (\emph{i.e.} the space needed for certification) and still have polynomial convergence time?
\end{open}
It is proved in \cite{BlinF15}, that this can be done for several variants of spanning trees.


\vspace{0.3cm}

\paragraph*{Conclusion}

In this paper, we introduced the domain of local certification, through its history and techniques. As we have seen in the last section, there is still a lot of open questions to answer and interesting directions to follow. We hope that this document will foster new research on the topic. 

\vspace{0.3cm}

\acknowledgements{
We would like to thank to the reviewers for their careful reading and suggestions that helped a lot in improving the paper.
Thanks to Fabien Dufoulon for several useful remarks, to Tatiana Starikovskaya for inviting me to give a seminar  which inspired Section~\ref{sec:directions}, to Ami Paz for pushing me to publish this document, and to the community of local certification for all the works in this domain. }


\DeclareUrlCommand{\Doi}{\urlstyle{same}}
\renewcommand{\doi}[1]{\href{https://doi.org/#1}{\footnotesize\sf doi:\Doi{#1}}}

\bibliographystyle{plainnat}
\bibliography{certif-biblio}
\end{document}